\documentstyle[12pt]{article}

\begin{document}

\tolerance=5000

\def\be{\begin{equation}}
\def\ee{\end{equation}}
\def\bea{\begin{eqnarray}}
\def\eea{\end{eqnarray}}
\def\nn{\nonumber \\}
\def\e{{\rm e}}

\def\SEH{S_{\rm EH}}
\def\SGH{S_{\rm GH}}
\def\AdS5{{{\rm AdS}_5}}
\def\S4{{{\rm S}_4}}

\  \hfill
\begin{minipage}{3.5cm}
December 2002 \\
\end{minipage}

\vfill

\begin{center}
{\large\bf Bulk and brane gauge propagator \\
on 5d AdS black hole }

\vfill

{\sc Shin'ichi NOJIRI}\footnote{nojiri@cc.nda.ac.jp}
and {\sc Sergei D. ODINTSOV}$^{\spadesuit}
$\footnote{
odintsov@mail.tomsknet.ru. Also at ICREA and IEEC, Barcelona, Spain. }

\vfill

{\sl Department of Applied Physics \\
National Defence Academy,
Hashirimizu Yokosuka 239-8686, JAPAN}

\vfill

{\sl $\spadesuit$
Lab. for Fundamental Studies,
Tomsk State Pedagogical University,
634041 Tomsk, RUSSIA}

\vfill

{\bf ABSTRACT}

\end{center}

The bulk gauge fields on 5d AdS black hole are discussed.
We construct the bulk (and the corresponding brane) gauge
propagator when black hole has large radius. The properties 
of gauge and ghost propagators are studied in both, minkowski
 or euclidean signature.
In euclidean formulation the propagator structure corresponds to the one of
 theory at finite temperature (which depends on coordinates).  
The decoupling of KK modes and localization 
of gauge fields on flat brane is demonstrated. We show that with such a bulk
there is no natural solution of hierarchy problem.

\vfill


\newpage

\noindent
1. Assuming that observable Universe is multi-dimensional one the fundamental
question appears: What is the form  (and number) of extra dimensions?
Are they planar or orbifolded?  Within the recent braneworld scenario
\cite{RS} the most likely possibility for Universe is 5d AdS space.
The braneworld scenario with bulk 5d AdS is very attractive 
as it may have the superstring origin.
Moreover, in such picture the four-dimensional world represents the brane 
(where graviton and matter fields are trapped) embedded into AdS bulk.
The nice way to solve the hierarchy problem on the brane appears.

Nevertheless, one can admit another choice for bulk manifold, for example,
AdS black hole. Indeed, like pure AdS space the AdS black hole may 
be (part of) superstring vacuum. The phase transition which may be
interpreted as confinement-deconfinement transition in AdS/CFT set-up may
occur
between pure AdS and AdS black hole\cite{witten}. Moreover, 
the graviton is localized on the brane embedded in AdS black hole\cite{bh}.
Hence, it is very interesting to understand the properties of such bulk choice
and their possible relation with the phenomenology of the unified theories.

In the present Letter we consider bulk abelian gauge fields in 5d AdS black
hole. The bulk (brane) gauge propagator is constructed for large black hole
with minkowskii or euclidean signature. 
The decoupling of KK modes and localization of gauge fields on flat brane 
is demonstrated. It is shown that unlike to pure AdS bulk there is no
natural solution for hierarchy problem. Brief discussion of possible phase
transition and its consequences for bulk QFT is given. 

\ 

\noindent
2. In order to consider the propagators, we start with the action 
of the abelian gauge theory in the 5-dimensional 
curved background:
\be
\label{I}
S_0=-{1 \over 4}\int d^5 x \sqrt{-g} g^{\mu\nu}g^{\rho\sigma}
F_{\mu\rho}F_{\nu\sigma}\ ,\quad 
F_{\mu\nu}=\partial_\mu A_\nu - \partial_\nu A_\mu\ .
\ee
The gauge fixing action $S_f$ and the ghost action 
$S_g$ are:
\be
\label{II}
S_f ={1 \over 2\xi}\int d^5 x \sqrt{-g}
\left(\nabla^\mu A_\mu\right)^2\ ,
\quad S_g=\int d^5 x \sqrt{-g} b \nabla^\mu \partial_\mu c\ .
\ee
Note that discussion of bulk gauge fields in  AdS braneworlds
has been given in refs.\cite{CH,HS,GP,AP,DH}.
As a curved background, we consider the Schwarzschild-Anti-de 
Sitter (SAdS) space with the flat horizon:
\be
\label{III}
ds^2=g_{\mu\nu}dx^\mu dx^\nu = - \e^{2\rho}dt^2 
+ \e^{-2\rho}dr^2 + r^2 \sum_{i=1,2,3}\left(dx^i\right)^2\ ,
\quad \e^{2\rho}={r^2 \over l^2} - {\mu \over r^2}\ .
\ee
In the following,  $\xi=-1$.  
The simplest choice corresponds to the large black hole,  that is, the
horizon 
radius $r_H=l^{1 \over 2}\mu^{1 \over 4}\to \infty$. We now 
define a new coordinate $s$ by $s=r-r_H$, then 
$\e^{2\rho}\sim {4 r_H \over l^2}s$, 
With new coordinate the action looks as 
\bea
\label{VI}
&& S \to \int d^5 x \left[ A_r\left\{ - {r_H^3 \over 2}
\partial_t^2 A_r + {2 r_H^2 \over l^2}s\triangle A_r
+ {8 r_H^5 \over l^4}s \partial_s^2 \left(sA_r\right)
\right\} \right. \nn
&& \ + A_t \left\{{r_H l^4 \over 32 s^2} \partial_t^2 A_t 
 - {r_H^3 \over 2}\partial_s^2 A_t - {l^2 \over 8s} 
 \triangle A_t\right\}  + {r_H^3 \over s}A_r \partial_t A_t 
+ A_i\left\{ - {l^2 \over 8s}\partial_t^2 A_i \right.\\
&& \left.\left. \ + {2 r_H^2 \over l^2}\partial_s\left(s\partial_s A_i\right) 
+ {1 \over 2r_H}\triangle A_i\right\} + b\left\{-{l^2 r_H^2 \over
4s}\partial_t^2 c 
 + r_H \triangle c + {4r_H^4 \over l^2}\partial_s\left(
 s\partial_s c\right)\right\}\right]\ .\nonumber
\eea
As we consider the plane waves in the $t$ and $x^i$ directions, 
we replace $\partial_t$ and $\triangle$ by $i\omega$ and 
$-k^2$, respectively. Then the equations of motion corresponding to 
the action (\ref{VI}) have the following form:
\bea
\label{VII}
&& 0= r_H^3 \omega^2 A_r - {4 r_H^2 k^2 \over l^2}s A_r 
+ {16 r_H^5 \over l^4}s \partial_s^2\left(sA_r\right) 
+ i{r_H^3 \omega \over s} A_t \ ,\\
\label{VIII}
&& 0 = - {r_H l^4 \omega^2 \over 16 s^2} A_t
 - r_H^3 \partial_s^2 A_t + {l^2 k^2 \over 4s}A_t 
 + i{r_H^3 \omega \over s}A_r \ , \\
\label{IX}
&& 0= {l^2 \omega^2 \over 8s}A_i 
+ {2r_H^2 \over l^2}\partial_s\left(s\partial_s A_i\right) 
 - {k^2 \over 2r_H}A_i\ ,\\
\label{X}
&&0= {l^2 \omega^2 \over 4r_H s} \psi - {k^2 \over r_H^2} \psi 
+ {4r_H \over l^2}s\partial_s\left(s\partial_s \psi\right)
\quad \psi=b,c\ .
\eea
We should note the equations (\ref{IX}) and (\ref{X}) for $A_i$, $c$ and $b$, 
respectively, are identical 
with each other.
If one defines a new complex field $B$ by $B\equiv sA_r + {il^2 \over
4r_H}A_t$, 
the equations (\ref{VII}) and (\ref{VIII}) can be combined to 
give
\be
\label{XIII}
0=\partial_s^2 B - {l^2 k^2 \over 4r_H^3 s}B 
+ {1 \over s^2}\left({l^4 \omega^2 \over 16 r_H^2}
+ {l^2 \omega \over 4 r_H}\right) B\ .
\ee
By further defining 
\be
\label{XIV}
B=s^{1 \over 2}\tilde B\ ,\quad s= {r_H^3 \over l^2 k^2} v^2\ ,
\ee
we can rewrite (\ref{XIII}) in the following form:
\be
\label{XV}
0={d^2 \tilde B \over dv^2} + {1 \over v}{d\tilde B \over dv} 
 - \left\{ 1 + {\nu_B^2 \over v^2}\right\}\tilde B\ ,\quad
\nu_B^2 \equiv 1 - {l^4 \omega^2 \over 4 r_H^2} 
 - {l^2 \omega \over r_H} \ ,
\ee
which is the modified Bessel's derivative equation. The 
solution is given by the linear combination of the modified 
Bessel functions $I_{\mu_B}(v)$ and $K_{\mu_B}'(v)$. Then the 
general solution for $B$ is given by
\be
\label{XVI}
B=s^{1 \over 2}\left[\alpha_0 I_{\nu_B} \left({lk s^{1 \over 2} 
\over r_H^{3 \over 2}}\right) + \beta_0 K_{\nu_B} 
\left({lk s^{1 \over 2} \over r_H^{3 \over 2}}\right)\right]\ .
\ee
Here $\alpha_0$ and $\beta_0$ are constants. 
On the other hand,  defining $v$ by (\ref{XIV}), 
Eq.(\ref{IX}) can be rewritten as
\be
\label{XVIII}
0={d^2 \tilde A_i \over dv^2} + {1 \over v}{d \tilde A_i \over dv} 
 - \left\{ 1 + {\nu_A^2 \over v^2}\right\}A_i\ ,
 \quad \nu_A^2 = - {l^4 \omega^2 \over 4 r_H^2}\ ,
\ee
whose solution is again given by the modified Bessel functions. 
In general, $\nu_A$ is imaginary. 
Then the general solution for $A_i$, and similarly the solutions for (\ref{X}) 
are given by (with constant $\alpha_{c,b}$ and $\beta_{c,b}$, $\psi=c,b$) 
\be
\label{XIX}
A_i=\alpha_i I_{\nu_A} \left({lk s^{1 \over 2} 
\over r_H^{3 \over 2}}\right) + \beta_i K_{\nu_A} 
\left({lk s^{1 \over 2} \over r_H^{3 \over 2}}\right),\  
\psi=\alpha_\psi I_{\nu_A} \left({lk s^{1 \over 2} 
\over r_H^{3 \over 2}}\right) + \beta_\psi K_{\nu_A} 
\left({lk s^{1 \over 2} \over r_H^{3 \over 2}}\right) .
\ee
Putting two branes at $s=s_1$ and $s=s_2$ ($0<s_1<s_2$) one 
glues two bulk spaces whose boundaries are the branes and imposes $Z_2$
symmetry. Then 
the gauge  and ghost fields should satisfy the following boundary
conditions
\be
\label{XXII}
\left.{\partial A_i \over \partial s}\right|_{s=s_{1,2}}=
\left.{\partial A_t \over \partial s}\right|_{s=s_{1,2}}=
\left.{\partial c \over \partial s}\right|_{s=s_{1,2}}=
\left.{\partial b \over \partial s}\right|_{s=s_{1,2}}=
\left.A_r\right|_{s=s_{1,2}}=0\ .
\ee
In general, the boundary conditions at $s=s_1$ are not 
consistent with those at $s=s_2$. We denote the fields 
which satisfy the boundary conditions at $s=s_a$ ($a=1,2$) 
by $A_{(a)\mu}$, $c_{(a)}$, $b_{(a)}$, etc. 
Using the boundary conditions (\ref{XXII}), the ratios
between the coefficients in (\ref{XVI}) and (\ref{XIX}) can be 
determined as follows:
\bea
\label{XXIII}
&& {\alpha_{(a)t} \over \beta_{(a)t}}
= {\left({1 \over 2}+ {\nu_B \over 2}\right)
K_{\nu_B}\left({lk s_a^{1 \over 2} \over r_H^{3 \over 2}}\right)
+ {lk s_a^{1 \over 2} \over 2r_H^{3 \over 2}}
K_{\nu_B-1}\left({lk s_a^{1 \over 2} \over r_H^{3 \over 2}}\right)
\over -\left({1 \over 2}+ {\nu_B \over 2}\right)
I_{\nu_B}\left({lk s_a^{1 \over 2} \over r_H^{3 \over 2}}\right)
+ {lk s_a^{1 \over 2} \over r_H^{3 \over 2}}
I_{\nu_B-1}\left({lk s_a^{1 \over 2} \over 
2 r_H^{3 \over 2}}\right)} \ ,\nn 
&& {\alpha_{(a)r} \over \beta_{(a)r}}=
 - {K_{\nu_B}\left({lk s_a^{1 \over 2} \over r_H^{3 \over 2}}\right)
\over I_{\nu_B}\left({lk s_a^{1 \over 2} \over r_H^{3 \over 2}}
\right)} \ ,\quad {\alpha_{(a)i} \over \beta_{(a)i}}
={\alpha_{(a)c} \over \beta_{(a)c}}
={\alpha_{(a)b} \over \beta_{(a)b}}
={\alpha_{(a)} \over \beta_{(a)}} \ ,\nn
&& \left\{\begin{array}{l}
\alpha_{(a)} =  {\nu_A \over 2}
K_{\nu_A}\left({lk s_a^{1 \over 2} \over r_H^{3 \over 2}}\right)
+ {lk s_a^{1 \over 2} \over 2 r_H^{3 \over 2}}
K_{\nu_A -1}\left({lk s_a^{1 \over 2} \over r_H^{3 \over 2}}\right) \\
\beta_{(a)} =  - {\nu_A \over 2}
I_{\nu_A}\left({lk s_a^{1 \over 2} \over r_H^{3 \over 2}}\right)
+ {lk s_a^{1 \over 2} \over 2 r_H^{3 \over 2}}
I_{\nu_A -1}\left({lk s_a^{1 \over 2} \over r_H^{3 \over 2}}\right) \\
\end{array}\right. \ .
\eea
Here  $\alpha_0$ and $\beta_0$ in (\ref{XVI}) are written as 
$\alpha_{(a)0}=\alpha_{(a)t} + i{l^2 \over 4r_H}\alpha_{(a)r}$, 
$\beta_{(a)0}=\beta_{(a)t} + i{l^2 \over 4r_H}\beta_{(a)r}$.

We now construct the propagators from the obtained classical 
solutions by following the procedure given in \cite{RSc}. 
One first constructs the propagator $G_{ij}$ 
for $A_i$:
\bea
\label{XXV}
G_{ij}(u,v)&=&N_1\delta_{ij}\left\{
A_{(2)}(u)A_{(1)}(v)\theta(u-v)
+ A_{(2)}(v)A_{(1)}(u)\theta(v-u)\right\} \ ,\nn
A_{(a)}(s)&\equiv & \alpha_{(a)} I_{\nu_A} \left({lk s^{1 \over 2} 
\over r_H^{3 \over 2}}\right) + \beta_{(a)} K_{\nu_A} 
\left({lk a^{1 \over 2} \over r_H^{3 \over 2}}\right)\ .
\eea
Here $N$ is a normalization constant determined later. 
We should note that the above propagator satisfies 
the boundary conditions, on both of the branes, 
corresponding to (\ref{XXII}):
$\left.{\partial G_{ij}(u,v) \over \partial u}\right|_{u=s_{1,2}}=
\left.{\partial G_{ij}(u,v) \over \partial v}\right|_{v=s_{1,2}}=0$.
As $A_{(a)}$ satisfies Eq.(\ref{IX}) 
one gets
\bea
\label{IXb}
&&{l^2 \omega^2 \over 8u}G_{ij}(u,v) 
+ {2r_H^2 \over l^2}\partial_u\left(u\partial_u G_{ij}(u,v)\right) 
 - {k^2 \over 2r_H}G_{ij}(u,v) \nn
&& = {2r_H^2 N \over l^2} u\left( {\partial A_{(2)}(u) \over 
\partial u}A_{(1)}(u) - A_{(2)}(u) {\partial A_{(1)}(u) \over 
\partial u} \right)\delta(u-v)\ .
\eea
 Then one can find the quantity 
$u\left( {\partial A_{(2)}(u) \over \partial u}A_{(1)}(u) 
 - A_{(2)}(u) {\partial A_{(1)}(u) \over \partial u} \right)$ does not
depend on $u$.
 We now can evaluate the quantity 
$u\left( {\partial A_{(2)}(u) \over \partial u}A_{(1)}(u) 
 - A_{(2)}(u) {\partial A_{(1)}(u) \over 
\partial u} \right)$ at $u=s_1$ by using the 
boundary condition (\ref{XXII}) and the expressions in 
(\ref{XXIII}):
\be
\label{XXXVIII}
u\left( {\partial A_{(2)}(u) \over \partial u}A_{(1)}(u) 
 - A_{(2)}(u) {\partial A_{(1)}(u) \over 
\partial u} \right) = -{1 \over 2}\left(\beta_{(1)} \alpha_{(2)} -
\alpha_{(1)}
\beta_{(2)}\right)\ .
\ee
In the last line, we have used a formula for the modified 
Bessel functions:
$I_\nu(z) K_{\nu - 1}(z) + I_{\nu - 1}(z) K_\nu (z) 
= {1 \over z}$. 
Since the r.h.s. of Eq.(\ref{IXb}) should be ${1 \over 2r_H^3}
\delta(u-v)$ since $\sqrt{-g}=r_H^3$, one finds the 
normalization constant should be 
$N=- {l^2 \over 2r_H^5
\left(\beta_{(1)} \alpha_{(2)} - \alpha_{(1)}
\beta_{(2)}\right)}$.

We now consider the correlation functions for $B$. 
>From the action (\ref{VI}), we find the correlator $G_{BB^*}(u,v)$ between $B$
and its complex conjugate vanishes. 
 One should be careful for the fact that the boundary 
condition for $A_r$ is different from $A_t$ as in (\ref{XXII}), 
we find the correlator between two $B$ is given by
\be
\label{IVXI}
G_{BB}(u,v)=-{l^4\left\{
B_{(2)}(u)B_{(1)}(v)\theta(u-v)
+ B_{(2)}(v)B_{(1)}(u)\theta(v-u)\right\} \over 4r_H^5 
\left(\beta_{0(1)} \alpha_{0(2)} - \alpha_{0(1)}
\beta_{0(2)}\right)}\ .
\ee
Denoting the correlators between two $A_t$'s, two 
$A_r$'s, $A_t$ and $A_r$ by $G_{tt}(u,v)$, $G_{rr}(u,v)$, 
$G_{tr}(u,v)=G_{rt}(v,u)$, respectively, one gets
\bea
\label{IVXII}
&& G_{BB}(u,v)= - {l^4 \over 16r_H^2}G_{tt}(u,v) 
+ uv G_{rr}(u,v) \nn
&& \ + {il^2 \over 4r_H}
\left(uG_{rt}(u,v) + vG_{tr}(u,v)\right)\ ,\nn
&& 0= G_{BB^*}(u,v) \nn
&& = {l^4 \over 16r_H^2}G_{tt}(u,v) 
+ uv G_{rr}(u,v) + {il^2 \over 4r_H}
\left(-uG_{rt}(u,v) + vG_{tr}(u,v)\right)\ .
\eea
Then 
\bea
\label{IVXIII}
&& G_{tt}(u,v)=-{8r_H^2 \over l^4}\Re G_{BB}(u,v)\ ,
\quad G_{rr}(u,v)={1 \over uv}\Re G_{BB}(u,v)\ ,\nn
&& uG_{rt}(u,v)=vG_{tr}(u,v)
={2r_H \over l^2}\Im G_{BB}(u,v)\ .
\eea
Here $\Re$ and $\Im$ express the real and imaginary parts, 
respectively. 

Let us investigate the behavior of the obtained Green 
functions on the branes. In order to investigate 
the Kaluza-Klein (KK) mode, we consider the case that 
$\omega=0$ and investigate where the poles of the Green 
functions exist with respect to $k^2$. We now consider 
$G_{ij}(u,v)$ with $u=v=s_1$ and $\omega=0$. 
Since $\omega=0$, $\nu_A$ in (\ref{XVIII}) vanishes. 
Since  the radius $r_H$ is large, we can use the
 asymptotic
expansion of the modified Bessel functions.
Then we find the following expression of $G_{ij}$
\be
\label{B2}
\left.G_{ij}(s_1,s_1)\right|_{\omega=0}\sim  \delta_{ij}{2 \over k^2 r_H^2}
{{1 \over 2} + {l^2 k^2 \over 8 r_H^3}\left(s_1 
+ s_2 \ln {s_2 \over s_1}\right) \over 
s_1 - s_2 + {l^2 k^2 \over 8r_H^3}\left(s_1^2 - s_2^2 
+ 2s_1s_2\ln {s_2 \over s_1}\right)}\ .
\ee
The propagator  (\ref{B2}) has a pole at 
\be
\label{B3}
k^2=0
\ee
and 
\be
\label{B4}
-k^2=m_1^2 \equiv {8r_H^3 \left(s_1 - s_2\right) \over 
l^2 \left(s_1^2 - s_2^2 + 2s_1 s_2 
\ln {s_2 \over s_1}\right)}\ .
\ee
The pole  (\ref{B3})  corresponds to the 
zero mode and that of $k^2=-m_1^2$  (\ref{B4}) to 
the first KK mode. 
When $k^2\sim 0$, the propagator behaves as
\be
\label{B5}
G_{ij}\sim {R_0 \delta_{ij} \over k^2}\ ,\quad 
R_0 \equiv -{1 \over r_H^2 \left(s_1 - s_2\right)}\ ,
\ee
and when $k^2\sim -m_1^2$
\be
\label{B5b}
G_{ij}\sim {R_1 \delta_{ij} \over k^2 + m_1^2}\ ,\quad 
R_1 \equiv -{1 -  {2\left(s_1 - s_2\right)\left(s_1 
+ s_2 \ln {s_2 \over s_1}\right)\over s_1^2 - s_2^2 
+ 2 s_1 s_2 \ln {s_2 \over s_1}}
\over r_H^2 \left(s_1 - s_2\right)}\ . 
\ee
$R_0$ and $R_1$  correspond to the square of  
 the wave functions of zero mode and the 
first KK mode, respectively. Then $R_0$ and $R_1$ 
express the coupling of these modes on the brane. 
Eq.(\ref{B5}) indicates that the coupling of the 
KK mode might not be small. Eq.(\ref{B4}) tells, however, the mass of 
the KK modes is very large since we are considering the large black 
hole ($r_H\to \infty$). Then the KK modes  decouple.

The above mass of the first KK mode (\ref{B4}) can be compared with 
the pure AdS case. By choosing the metric of the AdS as 
\be
\label{AdS1}
ds_{\rm AdS_5}^2={l^2 \over z^2}\left(dz^2 -dt^2 + \sum_{i=1,2,3}
\left(dx^i\right)^2\right)\ ,
\ee
we put two branes at $z=z_1=l$ and $z=z_2$. If $z_1<z_2$ and $1/z_2$ 
is the order of TeV, the brane at $z=z_2$ corresponds to the so-called 
TeV brane and that at $z=z_1$ to the Planck brane. Then the mass of the 
first KK mode is typically given by \cite{HS2}
\be
\label{AdS2}
\tilde m_1={a \over z_2 \sqrt{2\ln \left({z_2 \over z_1}\right)}} \ .
\ee
Here $a$ is a parameter describing the coupling of the gauge field to 
the TeV brane. Then if $a\sim 0.1$, the mass is weak scale. 
On the other hand, from (\ref{B4}), the mass corresponding to the first 
KK mode is given by
\be
\label{SAdS1}
\hat m_1^2={m_1^2 \over r_H^2}
= {8r_H \left(s_1 - s_2\right) \over 
l^2 \left(s_1^2 - s_2^2 + 2s_1 s_2 
\ln {s_2 \over s_1}\right)}\ .
\ee
The denominator $r_H^2$ in ${m_1^2 \over r_H^2}$ comes from the factor 
$r^2\sim r_H^2$ in front of $ \sum_{i=1,2,3}\left(dx^i\right)^2$ in the 
metric (\ref{III}). If $s_1\sim s_2\sim l$ and they are of the order of the 
Planck length, the mass $\hat m_1$ is 
$\hat m_1 \sim {1 \over l}\sqrt{r_H \over l}$.
Then the mass is much larger than the Planck mass for the large black hole.

It is interesting to understand  if the hierarchy problem with AdS BH 
can be solved in the same way as in \cite{RS}. 
Naively as scale factor is almost constant for the large black hole, it
would be 
difficult to realize the solution of hierarchy problem. 
Since $\e^{2\rho}\sim {4 r_H \over l^2}s$, the  metric looks as:
\be
\label{IIIB}
ds^2 = - {4r_H s \over l^2}dt^2 
+ {l^2 \over 4r_H s}ds^2 + r_H^2 \sum_{i=1,2,3}\left(dx^i\right)^2\ .
\ee
Then on the brane at $s=s_a$, if we redefine the coordinates by
$\tilde t = 2t\sqrt{s_a \over r_H}$, $\tilde x^i = l x^i$, 
the metric on the brane becomes flat
\be
\label{H2}
ds_{\rm brane}^2 = \sum_{m,n=0}^3\tilde g_{mn}dx^m dx^n
\equiv {r_H^2 \over l^2}\left( -d\tilde t^2 + \sum_{i=1,2,3}\left(d\tilde
x^i\right)^2\right) \ .
\ee
We now consider the gauge field $\tilde A_m$ and the Higgs fields $\phi$ 
 whose action on the brane is given by
\bea
\label{H3}
S_{\tilde A, \phi}
&=& \int d^4 x \sqrt{-\tilde g}\left[{1 \over 2} \left\{\tilde g^{mn} 
\left(\partial_m \phi^* + ie \tilde A_m \phi^*\right)
\left(\partial_n \phi -ie \tilde A_n \phi \right) \right.\right. \nn
&& \left.\left. - g^2 \left(\phi^* \phi - \lambda^2 \right)^2\right\}
 - {1 \over 4}\tilde g^{mn} \tilde g^{kl} F_{mk} F_{nl}\right] \nn
&=& \int d^4 x {r_H^4 \over l^4}\left[{1 \over 2} \left\{{l^2 \over r_H^2}
\eta^{mn} 
\left(\partial_m \phi^* + ie \tilde A_m \phi^*\right)
\left(\partial_n \phi -ie \tilde A_n \phi \right) \right.\right.\nn
&& \left.\left. - g^2 \left(\phi^* \phi - \lambda^2 \right)^2\right\}
 - {1 \over 4}{l^4 \over r_H^4}\eta^{mn} \eta^{kl} F_{mk} F_{nl}\right] \ .
\eea
Here $\eta_{mn}$ is the metric of the flat minkowski space. The couplings
$e$ and $g$ 
are of the order of the unity and $\lambda$ could be of the order of the
Planck 
scale, which gives $\lambda \sim {1 \over l}$. If we rescale the Higgs scalar 
field $\phi$ by $\phi={l \over r_H}\tilde \phi$, 
the action (\ref{H3}) can be rewritten by
\bea
\label{H5}
S_{\tilde A, \phi}
&=& \int d^4 x \left[{1 \over 2} \left\{\eta^{mn} 
\left(\partial_m \phi^* + ie \tilde A_m \phi^*\right)
\left(\partial_n \phi -ie \tilde A_n \phi \right)  \right.\right. \nn
&& \left.\left. - g^2 \left(\phi^* \phi - {\lambda^2 r_H^2 \over l^2}
\right)^2\right\}
 - {1 \over 4}\eta^{mn} \eta^{kl} F_{mk} F_{nl}\right] \ .
\eea
As the expectation value of $\left<\tilde \phi \right>\sim {\lambda r_H
\over l} 
\sim {r_H \over l^2}$ gives the mass of the gauge field $\tilde A_m$, in
order that 
$\left< \tilde \phi \right>\sim 10^2$ GeV, we have
${1 \over r_H}\sim 10^{36}\ \left(\mbox{GeV}\right)$, 
which may be unrealistic, and unfortunately may contradict with the
assumption that 
the black hole is large. The problem has occured because we consider the
brane outside 
the horizon where the warp factor becomes the large constant.

It is remarkable that the small $k$ behavior in (\ref{B5}) shows that the 
propagator behaves as $1/r$ in the position space, where $r$ is the 
distance between two points on the brane. The behavior is identical 
with usual propagator in four dimensions but not with that in 
five dimensions. Then (\ref{B5}) shows that the gauge fields are 
localized on the brane (for recent discussion of gauge fileds localization 
on brane embedded into pure AdS or dS bulks, see\cite{ghoroku}). 

So far the brane was considered near the horizon. 
One may consider the brane near the singularity, say $r\sim l$. Since we can  
regard the warp factor with $r^2$, the ratio of the scales on the brane and
on the 
brane near the horizon, where $r\sim \mu^{1 \over 4} l^{1 \over 2}$, is given 
by ${\mu^{1 \over 4} \over l^{1 \over 2}}$. Then  to reproduce the ratio of 
the Planck scale and the weak scale, we have ${\mu^{1 \over 4} \over l^{1
\over 2}}
\sim 10^{17}$. Since the mass $M$ of the black hole is given by 
$M\sim {\mu \over \kappa^2}$ and if $\kappa^2 \sim l^3$, we have 
$M\sim {\mu \over l^3} \sim 10^{68} {1 \over l}$. Since the Planck mass 
${1 \over l}$ is $\sim 10^{-5}$ gram, if the mass of the bulk black hole is 
macroscopic and given by $10^{63}$ gram $=10^{60}$ kg, the hierarchy might be 
realized.  Of course, such huge black hole mass seems to be unrealistic
and such effect may occur only if black hole quickly evaporate.
Another possibility which we mention in the discussion is that hierarchy
problem is solved with pure AdS bulk but then phase transition to
AdS black hole occurs.

\ 

\noindent
3. In this section, we Wick-rotate the metric (\ref{III}) into the 
Euclidean signature. Then since the black hole has (Hawking) temperature, 
one arrives at the finite temperature theory. Especially for the large 
black hole, the euclidean bulk spacetime surely becomes  flat. 

As a background bulk space,  the Schwarzschild-anti de Sitter space with 
the flat horizon is taken:
\be
\label{IIIb}
ds^2=g_{\mu\nu}dx^\mu dx^\nu = - \e^{2\rho}dt^2 + \e^{-2\rho}dr^2 
+ r^2 \sum_{i=1,2,3}\left(dx^i\right)^2\ ,
\quad \e^{2\rho}={r^2 \over l^2} - {\mu \over r^2}\ .
\ee
For the large black hole  the horizon 
radius $r_H=l^{1 \over 2}\mu^{1 \over 4}\to \infty$. We now 
define a new coordinate $\sigma$ by $\sigma=r-r_H$, then 
$\e^{2\rho}\sim {4 r_H \over l^2}\sigma$ and
\be
\label{LC1}
ds^2\to - {4 r_H \over l^2}\sigma dt^2 
+ {l^2 \over 4 r_H \sigma}d\sigma^2 
+ r_H^2 \sum_{i=1,2,3}\left(dx^i\right)^2\ .
\ee
After Wick-rotating time coordinate and  introducing 
new coordinate $\theta$ and $\rho$ by $\theta=i{2r_H \over l^2}t$,  
$\rho=l\sqrt{\sigma \over r_H}$, 
the metric (\ref{LC1}) can be rewritten as 
\be
\label{LC3}
ds^2 = \rho^2d\theta^2 + d\theta^2 
+ r_H^2 \sum_{i=1,2,3}\left(dx^i\right)^2\ .
\ee
In order to avoid the orbifold singularity at $\rho=0$, the 
coordinate $\theta$ has a period of $2\pi$ : 
$\theta\sim\theta + 2\pi$. Then the spacetime is flat and 
locally $R_2 \times R_3 \sim R^5$. 

We now further rescale the coordinates $x^i$ by $r_H x^i \to x^i$ and
 define a new field $B$ : $B\equiv A_\theta + i\rho A_\rho$. 
The part including only $B$ in the action has a $U(1)$ symmetry $B\to \e^{i\phi}B$ 
with constant parameter $\phi$ of the transformation.
As the coordinate $\theta$ has a period of $2\pi$, we can replace
$\partial_\theta$ by $in$. Here $n$ is an integer. Similarly, we replace 
$\partial_i$ by $k_i$ as the 
Fourier transformation. Then the solutions of the equations of motion are 
given by the modified Bessel functions with integer index $n$ by a way similar 
to the previous section. We now assume that there are branes at $\rho=\rho_1$, 
$\rho_2$ ($\rho_1<\rho_2$). One considers two classes of solutions, which 
satisfy the following boundary conditions at one of the two branes ($a=1,2$):
$\left.\partial_\rho A_{(a)i}\right|_{\rho=\rho_a}
=\left. A_{(a)\theta}\right|_{\rho=\rho_a}
=\left. A_{(a)\rho}\right|_{\rho=\rho_a} 
=\left.\partial_\rho c_{(a)}\right|_{\rho=\rho_a}
=\left.\partial_\rho b_{(a)}\right|_{\rho=\rho_a}
=0$. Here we have imposed the same boundary condition for 
$A_\theta$ and $A_\rho$ in order to preserve the $U(1)$ 
symmetry of $B$. Then we find 
\bea
\label{E11}
G_{ij}(u,v)&=& - {\delta_{ij} \over 2}
{A_{(2)}(u) A_{(1)}(v)\theta (u-v) + A_{(2)}(v) A_{(1)}(u)\theta (v-u)
\over \beta_{(1)}\alpha_{(2)} - \alpha_{(1)}\beta_{(2)}} \nn
G_{BB^*}(u,v)&=& - {1 \over 2}
{B_{(2)}(u) B^*_{(1)}(v)\theta (u-v) + B^*_{(2)}(v) B_{(1)}(u)\theta (v-u)
\over \beta_{(1)0}^*\alpha_{(2)0} - \alpha_{(1)0}\beta_{(2)0}^*} \ .
\eea
Here
\bea
\label{E12}
A_{(a)}&=& \alpha_{(a)} I_n(k\rho) + \beta_{(a)} K_n(k\rho) \ ,\nn
\alpha_{(a)} &=& -K_n'\left(k\rho_a\right) = K_{n-1}\left(k\rho_a\right) 
+ {n \over k\rho}K_n\left(k\rho_a\right) \ ,\nn
\beta_{(a)} &=& I_n'\left(k\rho_a\right)
= I_{n-1}\left(k\rho_a\right) - {n \over k\rho}I_n\left(k\rho_a\right) \ .
\eea
The coefficients are given by 
${\alpha_{(a)i} \over \beta_{(a)i}}={\alpha_{(a)c} \over \beta_{(a)c}}
={\alpha_{(a)b} \over \beta_{(a)b}} 
=-{K_n'\left(k\rho_a\right) \over I_n'\left(k\rho_a\right)} 
={K_{n-1}\left(k\rho_a\right) + {n \over k\rho}K_n\left(k\rho_a\right) 
\over I_{n-1}\left(k\rho_a\right) + {n \over k\rho}I_n\left(k\rho_a\right)}$, 
${\alpha_{(a)\theta} \over \beta_{(a)\theta}}
={\alpha_{(a)\rho} \over \beta_{(a)\rho}}
=-{K_{n-1}\left(k\rho_a\right) \over I_{n-1}\left(k\rho_a\right)}$. 
Here $\alpha_{(a)0}=\alpha_{(a)\theta} + i \alpha_{(a)\rho}$ and 
$\beta_{(a)0}=\beta_{(a)\theta} + i \beta_{(a)\rho}$.  
We also find the propagator between two $B$ or two $B^*$ vanishes.

We now consider the high energy behavior of the propagator 
when $k$, $n\to \infty$ but ${k \over n}$ is finite. 
Using asymptotic expansion of Bessel functions
one obtains
\bea
\label{EE2}
G_{ij}(u,v)&\sim& {\delta_{ij} \over 4|n|}
\left[{z_1^2 \over 1 + \sqrt{1+z_1^2}}
+ {z_2^2 \over 1 - \sqrt{1 + z_2^2}}\right]^{-1}
 {\sqrt{\left(1 + z_1^2\right)\left(1 + z_2^2\right)} 
\over \left(1+z_u^2\right)^{1 \over 4}
\left(1+z_v^2\right)^{1 \over 4}} \nn
&& \times \left\{\e^{|n|\left(\eta_v - \eta_u\right)}\theta(u-v)
+ \e^{|n|\left(\eta_u - \eta_v\right)}\theta(v-u)\right\} \ ,\nn
G_{BB^*}(u,v)&\sim & - {1 \over 4|n|}{\e^{|n-1|\left(\eta_v -
\eta_u\right)}\theta(u-v)
+ \e^{|n-1|\left(\eta_u - \eta_v\right)}\theta(v-u) \over 
\left(1+z_u^2\right)^{1 \over 4}\left(1+z_v^2\right)^{1 \over 4}} \ . 
\eea
Here 
\bea
\label{EE3}
&& z_a\equiv{k\rho_a \over n}\ ,\quad 
\eta_a\equiv \sqrt{1+z_a^2} + \ln {z_a \over 1 
+ \sqrt{1+z_a^2}} \ ,\nn
&& (a=1,2,u,v,\ \rho_u=u,\ \rho_v=v) \ .
\eea
 As a result, $G_{ij}(u,u)$, $G_{BB^* }(u,u)\propto 
{1 \over |n|\sqrt{1+z_u^2}} 
={1 \over u \sqrt{k^2 + {n^2 \over u^2}}}$. Then 
the behavior of the Green functions seems to be typical for theory at  
finite temperature. In fact, if we express the $D$-dimensional propagator 
at the finite temperature by $D-2$ spacial momenta $k_i$ ($i=1,2\cdots,D-2$) 
and 1 position $x^{D-1}$ and  further consider the case that $x^{D-1}=0$, 
we obtain
\be
\label{EE4b}
\hat G\left(k_1.k_2,\cdots,k_{D-2},x^{D-1}=0,n\right)
={\pi \over \sqrt{\sum_{i=1}^{D-1} k_i^2 + {n^2 \over \beta^2}}}\ ,
\ee
which corresponds to above brane propagator.
Here $\beta$ is the inverse of the temperature. 
We should also note that the brane temperature
$T={1 \over \beta}$ 
 seems to be $u$-dependent, $T\propto u$. Then for two branes, 
the inner brane is hotter than the outer brane. There will be a
thermal radiation 
from the inner brane to the outer brane. 

It is interesting that 
 there may appear the imaginary pole in the
propagator. 
Let us investigate the propagator $G_{ij}(u,v)$  (\ref{E11}). Taking $k=0$
 and extending $n$ to the complex continuum parameter, the propagator has a
pole at $n=0$, 
which is the usual massless pole. Besides the pole,  imaginary
poles appear, where the 
denominator of propagator  vanishes: 
$n=i{\pi N \over \ln {z_1 \over z_2}}$, 
where $N$ is an integer. 
It is known that imaginary part of the propagator is related with the
damping constant of the corresponding plasma state.
The fact that poles depend on the extra coordinates 
may lead to the following conjecture. The (part of) background radiation
(energy) may be due to the extra dimensions (if bulk is some black hole).
The effect should be universal and applied to any type of matter.
Hence, the temperature of hot matter may be presumably caused by 
the presence of bulk black hole.
This may provide the means for estimations of length of extra dimension(s).

\ 

\noindent
4. We presented the bulk and brane gauge propagator structure 
in AdS black hole bulk for both, euclidean and minkowski signatures 
of the metric. As it has been noted earlier for gravitons\cite{bh},
it is demonstrated that
there occurs localization of  gauge fields on the brane embedded into 5d
AdS black hole.  It is shown that KK modes decouple.
However, the solution of hierarchy problem when bulk is BH is 
not realistic.

Clearly, the study of gauge propagator is interesting in relation with 
possibility to apply it in diagrams for unified  theories.
Let us make several remarks in this connection.
We may introduce interaction  considering other bulk matter 
fields (scalars, spinors) or generalizing theory to non-abelian case. As one 
can see from (\ref{LC3}), the spacetime is locally flat although 
there are boundaries corresponding to the branes. Since the 
renormalization is determined by the short distance behavior, the 
beta-functions are not changed from their flat values. 
In more than four spacetime dimensions, the gauge coupling constant has
dimension. 
Then the beta-functions show power law behaviour except in four dimensions. 
If the spacetime dimension is less than four, the non-abelian gauge theory 
can be still asymptotic free but if the dimension is larger than four, 
there appears ultra-violet fixed point.
(The extra orbifold dimensions may completely change the structure 
of even trivial brane (scalar) theory and number of fixed points appear
or non-AF theory may become AF one\cite{milton}). Then we will have a
ultra-violet
fixed point in the 
Schwarzschild-AdS background with large black hole. 
For pure AdS, the gauge theories can be asymptotic free even in five
dimensions \cite{RSc, RG}. 
Then there might be a phase transition from asymptotic free theory to
asymptotic 
non-free theory at some critical horizon radius. 
(This may be another effect associated with AdS black hole,
similar to Hawking-Page-Witten phase transition \cite{witten}
interpreted as confinement-deconfinement phase transition via AdS/CFT.
This phase transition has also D-brane interpretation \cite{cvetic}.)
In case of the Schwarzschild black hole in the Minkowski background, the
larger 
black hole has the lower temperature but in case of Schwarzschild-AdS black 
hole with flat horizon, since the Hawking temperature is given by
$T_{H} = {r_H \over \pi l^2}$, 
the larger black hole has the higher temperature. Then the phase transition 
may be due to the temperature. The confinement would occur for asymptotic 
free gauge theories but would not occur for asymptotic non-free theories. 
Then the above phase transition, if it really exists, would correspond to 
the confinement-deconfinement phase transition. 

\ 

\noindent
{\bf Acknowledgments}
The work by S.N. is supported in part by the Ministry of Education, 
Science, Sports and Culture of Japan under the grant n. 13135208.

\end{document}